\documentclass[a4paper]{article}

\usepackage[english]{babel}
\usepackage[utf8x]{inputenc}
\usepackage{amsmath}
\usepackage{graphicx}
\usepackage[colorinlistoftodos]{todonotes}
\usepackage{authblk}

\title{A Dynamic Model of Functioning of a Bank.}

\author[1]{Oleg Malafeyev\thanks{malafeyevoa@mail.ru}}
\author[2]{Achal Awasthi\thanks{aa777@snu.edu.in}}
\affil[1]{Faculty of Applied Mathematics and Control Processes, Saint Petersburg State University, Russia}
\affil[2]{ Department Of Physics, Shiv Nadar University, India}
\date{\vspace{-5ex}}
\begin{document}
\maketitle

\begin{abstract}
In this paper, we analyze dynamic programming as a novel approach to solve the problem of maximizing the profits of a bank. The mathematical model of the problem and the description of a bank's work is described in this paper. The problem is then approached using the method of dynamic programming. Dynamic programming makes sure that the solutions obtained are globally optimal and numerically stable. The optimization process is set up as a discrete multi-stage decision process and solved with the help of dynamic programming.   
\end{abstract}

\section{Introduction and statement of the problem.}
\label{shaman}

We discuss about the problem of optimal functioning of a bank using a simple model. A bank's role is to grant loans and accept deposits. At each moment of time the bank has complete information about depositors and borrowers, i.e., the bank knows the volume of deposits, loans and their terms. The bank has the ability to calculate the borrowing and lending interest rates based in these data.
It is assumed that the lending (borrowing) interest rate, is a function that depends on the demand for loans (deposits), volumes of loans (deposits) and terms of the loan (Deposit). The bank reports the value of bets received for each client.

At the beginning of each period the bank receives loans and payments on deposits, after which it begins its considering the suggestions of investors.
The bank faces the challenge of maximizing profits by choosing an optimal policy, which also includes the depositors and
borrowers with whom the bank wants to work.
\section {Mathematical model of the problem.}

{\bf 1.\ The calculation of interest rates.}

Let us consider the following discrete model of the bank. There are $n$
periods and We will use the index
$i=1,\ldots,n$ to indicate the different periods. The bank operates on the time interval $[0,T]$.

At each moment of time $t_i:0=t_0<t_1< \ldots <t_n=T$, the bank
knows the number of borrowers $l_i$\negthinspace. The bank lends a known volume of capital $P{p_i}$ and decides the term $\tau _{p_i}$ (at which the borrower has to pay), for each of the borrower $p_i$ ($p_i=1,\ldots,l_i$). At any time, the bank may calculate the total amount of loanable capital in each period which is nothing but the demand for loans.It is given by:
$$
Q_{p_i}=\sum_{p_i=1}^{l_i}P_{p_i}.$$
To calculate interest rate on loans,a bank uses the following conditions:
\\
\noindent 1. An increase in the interest rate, increases the interest to be paid by the borrower. 

\noindent 2. An increase in demand for bank loans, increases the interest rates.

\noindent 3. Loan percentage and borrowing period are inversely related.
\\
\\
Let us assume that the interest rate at which money is lent to the borrower $p_i$
(let's denote $\eta_{\tau_{p_i}}$)
is a function that depends on
demand for
loans
$Q_{p_i}$, term period of the loan $\tau_{p_i}$ and the amount of being borrowed $P_{p_i}$
$$
\eta_{\tau_{p_i}}=\eta(Q_{p_i},P_{p_i},\tau_{p_i}).$$
Using the conditions described above, the function of the interest rate on loans can be represented in the following form:
$$\eta_{\tau_{p_i}}={1\over 100}a_{i} \exp{ \left [ {{P_{p_i}}\over {b_{1}}}-
{{\tau_{p_i}}\over {b_{2}}} \right ]},$$
where $b_{1}, b_{2}$ and $s$ are known and constant
coefficients, and the coefficient $a_i$ is calculated by the following recurrence relations:
$$a_i=a_{i-1}+{Q_{p_{i-1}}-Q_{p_{i-2}}\over s},$$
$a_0$ is the minimum borrowing interest rate which is known to us.
The borrowing interest rate $\rho_{\tau_{d_i}}$, the number of depositors $_{d_i}$, where $d_i=1,\ldots {m_i}$ ( where $m_i$ is the number of depositors in the current period) is a function of the contribution $D_{d_i}$ by the depositors $d_i$, the term $\tau_{d_i}$ and total deposits
$Q_{d_i}$ in period $i$.
Here,
$$ Q_{d_i}=\sum_{d_i=1}^{m_i}D_{d_i}.$$
The function of loan interest rate can be written as:-
$$\rho_{\tau_{d_i}}=\rho(Q_{d_i},D_{d_i},\tau_{d_i}).$$
The bank relies on the following conditions:

\noindent 1. An increase in the amount of loan, increases the debt percentage.

\noindent 2. Increasing the number of contributors, reduces the bank debt.

\noindent 3. The longer the implementation period, the higher the borrowed interest.

$$\rho_{\tau_{d_i}}={1\over 100}c_{i}\exp \left [{{D_{d_i}}\over {b_{1}}}+
{{\tau_{d_i}}\over {b_{2}}}\right] $$
The coefficients $b_1$, $b_2$ and $s$ are the same coefficients, which are used in the calculation of loan interest rates.
$$c_i=c_{i-1}+{Q_{d_{i-1}}-Q_{d_{i-2}}\over s}.$$
\\
The coefficients $c_0$ and $a_0$ satisfy the following condition: $a_0 \ge c_0$.
This is one way of calculating the
interest rates as already mentioned in the introduction. The bank can also offer loan by using a more complex procedure which includes using loan interest rates on demand of loans and deposits, amount of capital loaned and deposits, duration of loan and contribution.
\\
By counting loan amount and the borrowed interest, the bank has all the information required for beginning  its activity, i.e. the bank knows the amount of deposits and the amount loaned along with their terms, and interest rates.
\medskip
\\
{\bf 2.\ Description of the Bank's work.}
\\
On a finite time interval $[0,T]$ at discrete points $t_i:t_0=0<t_1<\ldots <t_n=T$, the bank has to choose clients with whom it is more profitable to work.
The following paragraph describes the sequence of operation of a bank. A bank, with initial capital $W_0$, decides to work with depositors
and borrowers. The bank stops its search when it chooses a particular client. 
We use the following assumption: the amount of money loaned in the first period should not exceed the current disposable capital in the bank. This is guided by the following constraint:
$$0 \le \sum_{p_1=1}^{l_1} \delta_{p_1}P_{p_1} \le W_0+ \sum_{d_1=1}^{m_1}
\Delta_{d_1}D_{d_1},$$
where the function $\Delta_{d_1}$ and $\delta_{p_1}$ are functions of favorable depositors and borrowers
$$\begin{array} {ccl}
\delta_{p_1} & = & \left \{ \begin{array}{ll} 1\,, &\mbox{if the bank lends to the borrower $P_{p_1}$,}\\
0\,, & \mbox{otherwise.}\\
\end{array} \right.\\
\\
\Delta_{d_1}& = & \left \{ \begin{array}{ll} 1\,, &\mbox{if the Bank uses the Deposit of $D_{d_1}$,}\\
0\,, & \mbox{otherwise.}\\
\end{array} \right.\\
\end{array}$$
{\bf Remark 1.} Based on the above assumption, we assume that during a period, the bank might not be able to lend money to all borrowers. 
(for example, if the available capital is less than the sum of volumes of all loans).
\\

At the end of the first period, at time $t_1$, the bank gets back loans with interest from the borrowers who have to repay their loans ($ \tau_{p_1}=1$) and has available funds. The bank makes payments on expired deposits, and stops its services for those customers, who the bank thinks at the expiration of the terms of the contribution
$ \tau_{d_1}=1,\ldots,n$ will not be able to pay the money due
the lack of capital available in the appropriate period ($i=1,\ldots,n$). Thus, the term deposits and loans equal each other,
$$\begin {array} {c} \displaystyle {W_0+ \sum_{d_1=1}^{m_1} \Delta_{d_1}D_{d_1}- \sum_{p_1=1}^{l_1} \delta_{p_1}P_{p_1}+ \sum_{p_1=1}^{l_1} \sum_{\tau_{{p_1}=1}} \delta_{p_1}P_{p_1}(\eta_{ \tau_{p_1}}+1) \ge}\\
\displaystyle { \ge \sum_{d_1=1}^{m_1}\sum_{\tau_{d_1}=1} \Delta_{d_1}D_{d_1}(\rho_{ \tau_{d_1}}+1).}\\
\end{array}$$
{\bf Remark 2.} In our model, it is assumed that the bank only has the ability to make the necessary payments as part of the contribution of one of its depositors and not take the entire amount.
\\
After completing these operations, at the beginning of the second period and end of the first period, i.e. at the end of $t_1$, the bank has the following amount of capital:
$$\begin {array} {c} \displaystyle {W(t_1)=W_0+ \sum_{d_1=1}^{m_1} \Delta_{d_1}D_{d_1}- \sum_{p_1=1}^{l_1} \delta_{p_1}P_{p_1}+}\\
\displaystyle{+ \sum_{p_1=1}^{l_1} \sum_{ \tau_{p_1}=1}\delta_{p_1}P_{p_1}(\eta_{ \tau_{p_1}}+1) -  \sum_{d_1=1}^{m_1} \sum_{ \tau_{d_1}=1} \Delta_{d_1}D_{d_1}(\rho_{ \tau_{d_1}}+1).}\\
\end{array}$$
or
\begin{equation}
\begin {array} {c} \displaystyle {W(t_1)=W_0+ \left [ \sum_{p_1=1}^{l_1} \sum_{ \tau_{p_1}=1} \delta_{p_1}P_{p_1} \eta_{ \tau_{p_1}}-  \sum_{d_1=1}^{m_1} \sum_{ \tau_{d_1}=1} \Delta_{d_1}D_{d_1} \rho_{ \tau_{d_1}} \right ]+}\\
\displaystyle {\qquad \qquad \qquad +\left [ \sum_{d_1=1}^{m_1} \sum_{ \tau_{d_1} \not = 1} \Delta_{d_1}D_{d_1}- \sum_{p_1=1}^{l_1}\sum_{ \tau_{p_1} \not = 1} \delta_{p_1}P_{p_1}\right ].}\\
\end{array}
\label{med_e24}
\end{equation}
The amount contained in the first bracket, refers to the bank's profit in the first step (period). We denote
$$F_1= \sum_{p_1=1}^{l_1} \sum_{ \tau_{p_1}=1} \delta_{p_1}P_{p_1}
\eta_{ \tau_{p_1}}-  \sum_{d_1=1}^{m_1} \sum_{ \tau_{d_1}=1}
\Delta_{d_1}D_{d_1} \rho_{ \tau_{d_1}}.$$
Before a bank starts work in the second period, it must make sure that $W(t_1) \ge W_0$. In subsequent periods, the bank functions in the same manner as in the first period.
\\
Consider a period $i$. The terms used by the bank when choosing its customers remain the same.
\\
\noindent
1. The amount of money that can be loaned in the $i$-th period shall not exceed the cash held in the bank at that point:
$$
0 \le \sum_{p_i=1}^{l_i} \delta_{p_i}P_{p_i} \le W(t_{i-1})+
\sum_{d_i=1}^{m_i} \Delta_{d_i}D_{d_i}.
$$
\noindent
2. The bank stops its services for those customers, who at the expiration of terms of the contribution, will not be able to pay the back the money due to the lack of capital:
$$
\begin {array} {c} \displaystyle {W(t_{i-1})+ \sum_{d_i=1}^{m_i} \Delta_{d_i}D_{d_i}- \sum_{p_i=1}^{l_i} \delta_{p_i}P_{p_i}+\sum_{j=1}^i \sum_{p_j=1}^{l_j} \sum_{ \tau_{p_j}=i-j+1} \delta_{p_j}P_{p_j}(\eta_{ \tau_{p_j}}+1) \ge} \\
 \displaystyle {\ge \sum_{j=1}^i \sum_{d_j=1}^{m_j} \sum_{ \tau_{d_j}=i-j+1}  \Delta_{d_j}D_{d_j}(\rho_{ \tau_{d_j}}+1).}
\end{array}
$$
\noindent 3. A bank should always have some amount of money in liquid form.
$$W(t_i)>0.$$
\noindent 4.A bank shouldn't operate in a loss:
$$W(t_i) \ge W_0.$$
Here,
$$\begin{array} {ccl}
\delta_{p_i} & = & \left \{ \begin{array}{ll} 1\,, &\mbox{if the bank lends to the borrower $P_{p_i}$,}\\
0\,, & \mbox{otherwise.}\\
\end{array} \right.\\
\\
\Delta_{d_i}& = & \left \{ \begin{array}{ll} 1\,, &\mbox{if the bank uses the deposit of $D_{d_i}$,}\\
0\,, & \mbox{otherwise.}\\
\end{array} \right.\\
\end{array}.$$
Similarly, formula (\ref{med_e24}) describes the amount of capital available to the bank at the end
$i$-th period:
$$
\begin{array} {r} \displaystyle {W(t_i)=W(t_{i-1})+ \left [\sum_{j=1}^i \sum_{d_j=1}^{m_j} \sum_{ \tau_{d_j} \not = i-j+1} \Delta_{d_j}D_{d_j} -\sum_{j=1}^i \sum_{p_j=1}^{l_j} \sum_{ \tau_{p_j} \not = i-j+1} \delta_{p_j}P_{p_j} \right ]+ }\\
\displaystyle{+\left [ \sum_{j=1}^i \sum_{p_j=1}^{l_j} \sum_{ \tau_{p_j}=i-j+1} \delta_{p_j}P_{p_j} \eta_{ \tau_{p_j}} - \sum_{j=1}^{i} \sum_{d_j=1}^{m_j} \sum_{ \tau_{d_j}=i-j+1} \Delta_{d_j}D_{d_j} \rho_{\tau_{d_j}} \right ].} \quad \\
\end{array}
$$
We denote,
$$F_i= \sum_{j=1}^i \left [ \sum_{p_j=1}^{l_j} \sum_{ \tau_{p_j}=i-j+1}
\delta_{p_j}P_{p_j} \eta_{ \tau_{p_j}}- \sum_{d_j=1}^{m_j} \sum_{
\tau_{d_j}=i-j+1} \Delta_{d_j}D_{d_j} \rho_{ \tau_{d_j}} \right]$$
as the total profit function for $i$ steps.
$$A_i= \sum_{j=1}^i \left [\sum_{d_j=1}^{m_j} \sum_{ \tau_{d_j} \not = i-j+1} \Delta_{d_j}D_{d_j}-\sum_{p_j=1}^{l_j} \sum_{ \tau_{p_j} \not = i-j+1} \delta_{p_j}P_{p_j} \right]. $$
Here, $A_i$ is the remaining or balance capital.
Then, using these designations, the cash capital will be of the form:
$$W(t_i)=W_0+F_i+A_i.$$
At the end of the period, a bank announces the results of its work that denotes that the balance capital should be equal to zero, i.e.
$$W(t_n)=W_0+F_n.$$
Thus, the total income of the bank at time $t_n=T$ will be equal to:
$$F(t_n)=F_n.$$
This is the value which is important to the bank and the bank must maximize this by choosing feasible
customers throughout the entire period.

\subsection{Description of the problem in terms of dynamic programming.}

{\bf 1. Deterministic case.}
\\
We will use
dynamic programming as a tool to optimize the bank's profits. Dynamic programming is based on the principle optimality: optimal behavior has the property that whatever
the initial condition and the solution at the initial time be, the subsequent decisions should be an optimal behavior of the state,
resulting from the first solution.
\\
Let us consider the case when there is no uncertainty, i.e. all the data
objectives are precise.
\\
Let us consider that the initial amount of capital the bank $W_0$ as
the initial state of the system. The value $W_0, W(t_1), \ldots,W(t_n)$
represent the amount of capital in the respective periods, which is
determined by the following formula:
$$ \begin{array} {r} \displaystyle {W(t_i)=W(t_{i-1})+ \left [\sum_{j=1}^i \sum_{d_j=1}^{m_j} \sum_{ \tau_{d_j} \not = i-j+1} \Delta_{d_j}D_{d_j} -\sum_{j=1}^i \sum_{p_j=1}^{l_j} \sum_{ \tau_{p_j} \not = i-j+1} \delta_{p_j}P_{p_j} \right ]+ }\\
\displaystyle{+\left [ \sum_{j=1}^i \sum_{p_j=1}^{l_j} \sum_{
\tau_{p_j}=i-j+1} \delta_{p_j}P_{p_j} \eta_{ \tau_{p_j}} -
\sum_{j=1}^{i} \sum_{d_j=1}^{m_j} \sum_{ \tau_{d_j}=i-j+1}
\Delta_{d_j}D_{d_j} \rho_{\tau_{d_j}} \right ],}
\end{array} $$
The efficiency of the entire process is determined by the income received by the bank within $n$ periods, described by the formula:
\begin{equation}
R_n(W_{0},W({t_1}),...,W({t_n}))=W(t_n)-W_0. \label{med_eq1}
\end{equation}
Now, we have to solve solving the problem of maximizing function (\ref{med_eq1}). Rather than considering one problem with $W(t_i)$ amount of capital and a fixed number of processes, we consider
a whole family of such problems, in which $W$ can take any positive value and $n$ can take any integer value.
Let us introduce the sequence of functions $\{f_n(W_0)\}$, which are defined for $n=2,\ldots,N,\ W_0>0$, as follows:
$$f_n(W_0)= \max_{\{q_i\}}{ R_n(W_{0},W({t_1}),...,W({t_n}))},$$
where $q_i=({\delta_{p_i},\Delta_{d_i}})$ is the  selection policy of depositors and borrowers of the bank (These are the control variables).
\\
The function $f_n(W_0)$ represents the optimal profit obtained from a $n$-step process that starts with $W_0 \ge 0$ capital. 
\\
For the first period, the optimal policy will have the following functional form:
\begin{equation}
f_1(W_0)= \max_{{q_1}\in{Q_1}} \left[ \sum_{p_1=1}^{l_1} \sum_{
\tau_{p_1}=1} \delta_{p_1}P_{p_1} \eta_{ \tau_{p_1}}-
\sum_{d_1=1}^{m_1} \sum_{ \tau_{d_1}=1} \Delta_{d_1}D_{d_1} \rho_{
\tau_{d_1}}\right]. \label{med2.1.3}
\end{equation}
where $Q_1$ is the set of admissible controls of a one-step process which are described by the following constraints:
$$\left \{ \begin{array}{l}
  \displaystyle { \sum_{p_1=1}^{l_1} \sum_{\tau_{p_1}=1} \delta_{p_1}P_{p_1} \le W_0+ \sum_{d_1=1}^{m_1}\sum_{\tau_{d_1}=1} \Delta_{d_1}D_{d_1}, }\\
   \\
  \begin {array} {c} \displaystyle {W_0+ \sum_{d_1=1}^{m_1} \sum_{\tau_{d_1=1}}\Delta_{d_1}D_{d_1}- \sum_{p_1=1}^{l_1} \sum_{\tau_{p_1}}\delta_{p_1}P_{p_1}+ \sum_{p_1=1}^{l_1} \sum_{ \tau_{p_1}=1} \delta_{p_1}P_{p_1}(\eta_{ \tau_{p_1}}+1) \ge}\\
\displaystyle { \ge \sum_{d_1=1}^{m_1}\sum_{\tau_{d_1}=1} \Delta_{d_1}D_{d_1}(\rho_{ \tau_{d_1}}+1).}\\
\end{array}\\
   \\
\begin{array}{ccl}\delta_{p_1} & = & \left \{ \begin{array}{ll} 1\,, &\mbox{if the bank lends to the borrower $P_{p_1}$,}\\
0\,, & \mbox{otherwise}\\
\end{array} \right.\\
\\
\Delta_{d_1}& = & \left \{ \begin{array}{ll} 1\,, &\mbox{if the bank uses the Deposit $D_{d_1}$,}\\
0\,, & \mbox{otherwise}\\
\end{array} \right.\\
\end{array}
\\
\begin {array} {l} \displaystyle {W(t_1)=W_0+ \sum_{d_1=1}^{m_1}\sum_{\tau_{d_1}} \Delta_{d_1}D_{d_1}- \sum_{p_1=1}^{l_1}\sum_{\tau_{p_1}} \delta_{p_1}P_{p_1}+}\\
\displaystyle{+ \sum_{p_1=1}^{l_1} \sum_{ \tau_{p_1}=1}\delta_{p_1}P_{p_1}(\eta_{ \tau_{p_1}}+1) -  \sum_{d_1=1}^{m_1}\sum_{ \tau_{d_1}=1} \Delta_{d_1}D_{d_1}(\rho_{ \tau_{d_1}}+1) \ge W_0.}\\
\end{array}
\end{array}
\right. $$
In a two-step process, the total income will consist of income from the first step plus income from the second step.
This explicitly states that the initially selected value of $W_0$, which has now become $W(t_1)$ must be used in the most advantageous manner. If $q_1=({\delta_{p_1},\Delta_{d_1}})$ is optimal, then from the initial distribution and from the second step of our two-step process, we get the full income $f_{1}{(W_1)}$.
Therefore, the final income from a two-step process when the initial distributional amount is $W_0$ is obtained by the following expression:
$$R_{2}(W_0,W(t_1))=\sum_{p_1=1}^{l_1} \sum_{ \tau_{p_1}=1} \delta_{p_1}P_{p_1} \eta_{ \tau_{p_1}}-  \sum_{d_1=1}^{m_1} \sum_{ \tau_{d_1}=1} \Delta_{d_1}D_{d_1} \rho_{ \tau_{d_1}}+f_{1}(W{(t_1)}).$$
Accordingly, a recurrence relation
$$f_2(W_0)= \max_{{q_1}\in{Q_1}} \left[ \sum_{p_1=1}^{l_1} \sum_{ \tau_{p_1}=1}
\delta_{p_1}P_{p_1} \eta_{ \tau_{p_1}}-  \sum_{d_1=1}^{m_1} \sum_{
\tau_{d_1}=1} \Delta_{d_1}D_{d_1} \rho_{
\tau_{d_1}}+f_{1}(W({t_1}))\right],$$
links functions $f_1(W_0)$ and $f_2(W_0)$. Using same
same the argument for a $n$-step process, we will obtain a functional
equation of the following form:
$$f_n(W_0)= \max_{{q_1}\in{Q_1}} \left[ \sum_{p_1=1}^{l_1} \sum_{ \tau_{p_1}=1}
\delta_{p_1}P_{p_1} \eta_{ \tau_{p_1}}-  \sum_{d_1=1}^{m_1} \sum_{
\tau_{d_1}=1} \Delta_{d_1}D_{d_1} \rho_{
\tau_{d_1}}+f_{n-1}(W({t_1}))\right].$$
for $n \ge 2$, where $f_1(W_0)$ is determined by the ratio
(\ref{med2.1.3}).
\\
\section{Conclusion}
Thus, if we know the value initial capital and by applying the method of dynamic programming we would get a sequence of functions $\{f_i(W_0)\}$ which corresponds to the function of maximum income, and the corresponding vector functions $\{q_i=(\delta_{p_i}, \Delta_{d_i})\}$ -- is optimal management at each step.

\section{References}
1)	Jong Min Lee, Jay H. Lee (2004), Approximate Dynamic Programming Strategies and Their Applicability for Process Control: A Review and Future Directions, International Journal of Control, Automation, and Systems, vol. 2, no. 3, pp. 263-278. 
\\
2)	R. Neuneier (1997), “Enhancing Q-learning for optimal asset allocation,” In M. Jordan, M. Kearns, and S. Solla, editors, Advances in Neural Information Processing Systems, vol. 10.
\\
3)	Xeniya Grigorieva, Oleg Malafeev A Competitive Many-period Postman Problem With Varying Parameters, Applied Mathematical Sciences, vol. 8, 2014, no. 146, 7249 – 7258.
\\
4)	Malafeyev, O.A., Neverova, E.G.,Nemnyugin, S.A., Alferov, G.V. Multi-criteria model of laser radiation control, Emission Electronics (ICEE), 2014 2nd International Conference on June 30 2014-July 4 2014, DOI: 10.1109/Emission.2014.6893966, Publisher: IEEE
\\
5)	Malafeev, O.A. Stationary strategies in differential games, USSR Computational Mathematics and Mathematical Physics Volume 17, Issue 1, 1977, Pages 37–46.
\\
6)	Malafeev O.A. Equilibrium situations in dynamic games, Cybernetics and System Analisys May–June, 1974, Volume 10, Issue 3, pp 504-513.
\\
7)	Edwin J. Elton, Martin J. Gruber (1971), Dynamic Programming Applications in Finance, The Journal Of Finance, Volume 26, Issue 2, Papers and Proceedings of the Twenty-Sixth Annual Meeting of the American Finance Association Detroit, Michigan, December 28-30, pp. 473-506. 
\\
8)	Malafeyev, O.A., Redinskikh, N.D., Alferov, G.V. Electric circuits analogies in economics modeling: Corruption networks, Emission Electronics (ICEE), 2014 2nd International Conference on June 30 2014-July 4 2014, DOI: 10.1109/Emission.2014.6893965, Publisher: IEEE.
\\
9)	Malafeyev, O.A., Nemnyugin, S.A., Alferov, G.V. Charged particles beam focusing with uncontrollable changing parameters, Emission Electronics (ICEE), 2014 2nd International Conference on June 30 2014-July 4 2014, DOI: 10.1109/Emission.2014.6893964, Publisher: IEEE.
\\
10)	Malafeev, O.A., Nemnyugin, S.A. Generalized dynamic model of a system moving in an external field with stochastic components, Theoretical and Mathematical Physics (Impact Factor: 0.7). 01/1996; 107(3):770-774. DOI: 10.1007/BF02070384.
\\
11)	Richard Bellman (1957), Dynamic Programming (Princeton University Press; Princeton NJ). 
\\
12)	William Beranek (1963), Analysis for Financial Decisions, (R.D. Irwin; Homewood).

\end{document}